\documentstyle[12pt]{article}
\begin{document}
\thispagestyle{empty}
\begin{flushright}
{\sc BU-HEP} 92-04\\January 1992
\end{flushright}
\vspace{.1cm}
\begin{center}
{\LARGE\sc Baryogenesis Constraints on the Minimal\\[10mm]
Supersymmetric Model\footnote{This work was supported in part
under NSF contract PHY--9057173, under DOE contracts
DE--FG02--91ER40676 ,
DE--AC02--89ER40509 and by funds from the Texas National Research
Laboratory Commission under grant RGFY91B6.}}\\[2cm]
{\sc Stanley Myint}\\[1.5cm]
{\sc Department of Physics\\[3mm]
Boston University\\[3mm]
Boston, MA 02215, USA}\\[1.5cm]
{\em to appear in Phys. Lett. B}\\[1.5cm]
{\sc Abstract}
\end{center}
\begin{quote}
Requirement that the vacuum expectation values
of Higgs fields immediately after the
phase transition be large enough imposes constraints upon the
parameters of the minimal supersymmetric model. In particular,
one obtains the upper bounds on the lighter CP-even Higgs mass and
the soft supersymmetry breaking scale for different values of the
top quark mass.
\end{quote}
\vfill
\newpage

\section{Introduction}
Several recent papers have studied the possibility of baryogenesis in
the Standard Model and its minimal extensions. For example, references
[1] - [6] impose constraints on the parameters of the Standard Model,
minimal supersymmetric model (MSUSY), models with additional bosons or
singlet Majoron model. In particular, authors of ref. [1] estimate the
upper limit on the lighter CP-even Higgs in MSUSY to be equal to that in
the Standard Model which they calculate to be 55 GeV. However, they have
restricted their analysis to the case of the top quark lighter than
about 115 GeV. The present paper includes the corrections due to the
heavy top and its supersymmetric partners.

The obvious question is: are these corrections important? Contributions
of the heavy top and stop to the effective potential have been
calculated in many papers ( [7] - [13] ) and turn out to be a
{\em dominant part of the one loop effective potential} . These results
have been used in ref. [14] in order to impose limits on the parameters
of MSUSY from existing data at LEP and in ref. [15] to analyze Higgs
signals in the future hadron supercolliders.

In this paper we consider the one loop effective potential including
contributions from W and Z gauge bosons, top and stop quarks and two
CP-even Higgs bosons of the MSUSY in the region of parameter space
where they play a non-negligible role.

Next we numerically calculate the zero and nonzero temperature VEV's
and impose the requirement that the latter be large enough immediately
after the phase transition to sufficiently suppress baryon number
violating processes. This constraint
gives upper bounds on the lighter CP-even Higgs mass and on
the soft SUSY breaking scale $\mu$.

In section 2, we briefly review how baryogenesis imposes constraints on
parameters of the Standard Model and in section 3 we give a short
description of the Higgs sector of MSUSY. Our calculation is
presented in section 4 and the results are analyzed in section 5.

\section{Constraints from Electroweak Baryogenesis}

Much attention has been devoted recently to the possibility of creating
the baryon asymmetry during the electroweak phase transition. For this
to be true one has to satisfy certain basic requirements. First, the
amount of CP violation inherent in the model has to be large enough to
account for the observed baryon to photon ratio of about $10^{-8}$.
Standard Model does not satisfy this requirement but it is possible to
extend it to have enough CP violation. Secondly, the phase transition
has to be first order to provide for departure from thermal equilibrium.

Once these requirements are met it is possible to construct a mechanism
which shows how the baryon asymmetry was created during the electroweak
phase transition, ( see for example ref. [3] ) instead of at much higher
energies. This lower energy scale will allow for predictions of
baryogenesis to be experimentally verifiable.

Regardless of the details of the particular mechanism of creating the
baryon asymmetry, it is important to make sure that baryon number
violating processes after the phase transition do not erase any
previously created symmetry. To be more specific: after the phase
transition at temperature $T_c$, the Higgs field acquires a vacuum
expectation value $v(T_c)$. The rate of baryon number violation by
thermal fluctuations is proportional to the Boltzmann factor
$exp(-M_{sph}/T_c)$, where $M_{sph}$ is the mass of the sphaleron
field configuration or equivalently the height of the barrier
separating the gauge field configurations with different baryon
numbers. $M_{sph}$ was calculated at zero $T$ in ref.[16]:
\begin{equation}
M_{sph}(T_c)= 4\pi B(\frac{\lambda}{g^2})\frac{v}{g_w}
\end{equation}
with: $B(\frac{\lambda}{g^2})\in [1.52,2.70]$, for $\frac{\lambda}{g^2}
\in [0,\infty) $, where the $\lambda$ is Higgs self coupling.

These baryon number violating processes have to proceed at a rate much
smaller than the expansion rate of the universe after the transition,
therefore the exponent of the Boltzmann factor has to be sufficiently
large. Shaposhnikov [17] showed that:

\begin{equation}
\frac{M_{sph}(T_c)}{T_c} \geq 45
\end{equation}

This equation was derived under the assumption [22] that at $T=T_c$
(1) can be approximated by the same expression with $v, \lambda, g^2$
evaluated at $T=T_c$ .

How does this constrain parameters of the model?
For instance, in the Standard Model, the one loop effective potential
$V^T(\phi)$ of the Higgs field at a temperature T much higher than the
masses of the particles is given by:

\begin{equation}
V^T(\phi)=\gamma(T^2-T^2_2) \phi^2-E T \phi^3+\frac{\lambda_T}
{4} \phi^4
\end{equation}
where the coefficients $\gamma ,E,\lambda _T$ are positive and are
determined by the parameters of the Standard Model. In particular,
$\lambda _T$ is the temperature dependent effective quartic self
coupling of the Higgs field.

The phase transition occurs near the point where the temperature
dependent effective mass of the Higgs field vanishes:
\begin{equation}
\gamma (T^2-T^2_2) \approx 0
\end{equation}

Then, the VEV is given by:
\begin{equation}
v(T_c)\approx \frac {3 E T} {\lambda _T}
\end{equation}

If we want to have $v(T_c)$ sufficiently large to satisfy (1) and
(2), we must have $\lambda_T$ sufficiently small. This in turn
imposes the upper limit on the Higgs self coupling $\lambda$ and
therefore on the Higgs mass.

The similar thing will happen in MSUSY as our results will show.
\section{Higgs Sector of MSUSY}
Higgs sector of MSUSY (see for instance ref. [18]) contains two
complex Higgs doublets with the following
$SU(3)\times SU(2)\times U(1)$ quantum numbers:
\begin{equation}
H_1=\left( \begin{array}{c}
        H^0_1 \\ H^-_1
          \end{array} \right) \in (1,2,-1/2)~,~
H_2=\left( \begin{array}{c}
        H^+_2 \\ H^0_2
          \end{array} \right) \in (1,2,+1/2)
\end{equation}

{}From these eight real fields spontaneous symmetry breaking decouples
three unphysical Goldstone bosons and one is left with five physical
Higgs bosons, namely: two CP-even scalars, one CP-odd scalar and
a pair of charged scalars.
{\samepage
The tree level Higgs potential:
\begin{eqnarray}
V & = & m^2_1 |H_1|^2 + m^2_2 |H_2|^2 - m^2_3 (H_1 H_2 + h.c.)
+\frac{g^2_1}{8} (H^+_1 \vec{\sigma} H_1 +H^+_2 \vec{\sigma} H_2)^2
\nonumber \\
  &   & + \frac{g^2_2}{8} (|H_1|^2 -|H_2|^2 )^2
\end{eqnarray}
}
can be restricted to the real components of the neutral Higgs fields,
$\phi_1~=~Re~H^0_1$, $\phi_2~=~Re~H^0_2$:
\begin{equation}
V_{tree}=m^2_1 |\phi_1|^2 + m^2_2 |\phi_2|^2 - m^2_3 \phi_1 \phi_2
         +\frac {g^2_1+g^2_2}{8} (\phi^2_1 - \phi^2_2)^2
\end{equation}

One can always choose such a field basis that $m_3 ^2$, $v_1$, $v_2$ are
real and positive. Constants $m_1$, $m_2$ and $m_3$ have to satisfy
certain conditions. Requiring that the potential be bounded from below
gives:
\begin{equation}
\frac{m^2_1+m^2_2}{2} \geq m^2_3
\end{equation}
and the spontaneous symmetry breaking condition is:
\begin{equation}
m^4_3 \geq m^2_1 m^2_2
\end{equation}
Here $v_1$ and $v_2$ are proportional to vacuum expectation values of
$\phi _1$ and $\phi _2$:
\begin{equation}
<\phi_1>\equiv \frac{v_1}{\sqrt{2}},
<\phi_2>\equiv \frac{v_2}{\sqrt{2}}
\end{equation}
and $tan\beta $ is defined to be their ratio: $tan\beta \equiv v2/v1$.
Here:
\begin{equation}
\sqrt{v^2_1+v^2_2}=246~GeV
\end{equation}
to reproduce the measured values of gauge boson masses:
\begin{equation}
m^2_w=\frac{g^2_1}{4}(v^2_1+v^2_2)~,~
m^2_z=\frac{g^2_1+g^2_2}{4}(v^2_1+v^2_2)
\end{equation}
Fields $\phi _1$ and $\phi _2$ couple to down and up type quarks
respectively. For example, after the spontaneous symmetry breaking
top quark gets the mass:
\begin{equation}
m^2_t=h^2_t<\phi_2>^2
\end{equation}

However, if the supersymmetry is softly broken, its scalar superpartner
``stop'' will have the different mass:
\begin{equation}
m^2_{\tilde{t}}=h^2_t<\phi_2>^2 +\mu^2
\end{equation}
where we consider only the case of common soft supersymmetry-breaking
mass $\mu $ for $\tilde{t}_L$ and $\tilde{t}_R$
and vanishing off-diagonal
elements of the $2\times 2$ stop mass matrix. (The following
analysis can be easily generalized to include all of these terms.)

Finally CP-even physical eigenstates h and H, with masses $m_h <
m_H$ are obtained by diagonalizing the mass matrix:
\begin{equation}
{\cal M} \equiv \frac {1}{2} \left( \frac {\partial^2 V_{tree}}
{\partial \phi_i \partial \phi_j} \right)_{min}=
\left( \begin{array}{cc}
        A & C \\
        C & B
\end{array}
\right)
\end{equation}
and their masses are given by:
\begin{equation}
m^2_{h,H}= \frac{1}{2}(A+B\mp\sqrt{(A-B)^2+4 C^2})
\end{equation}

All this was at the tree level, but, as was already mentioned, in
order to obtain limits on the Higgs mass it is paramount to include one
-loop corrections.
In the effective potential approach used in ref. [7] and [8]
, masses of the Higgs bosons are approximated with the eigenvalues of
the matrix of second derivatives of the one-loop effective potential
evaluated at its minimum.

The one loop effective potential at zero temperature
is given by the expression:
\begin{equation}
V^0=V_{tree}(Q)+\frac{1}{64 \pi^2} {\cal S}tr
           \left\{ {\cal M}^4(\phi) \left[ \ln \frac
           { {\cal M}^2(\phi)}{Q^2}-\frac{3}{2} \right] \right\}
\end{equation}

Here ${\cal M}^2(\phi)$is the field dependent squared mass
matrix, $Q$ is the renormalization scale and
the supertrace is given by:
\begin{equation}
{\cal S}tr~ f\left( {\cal M}^2 \right) = \sum_i (-1)^{2 J_i} g_i
                                     f \left( m^2_i \right)
\end{equation}

The sum runs over all the physical particles i of spin $J_i$, field
dependent mass eigenvalue $m_i$ and multiplicity $g_i$ that couple
to fields $\phi _1$ and $\phi _2$. In our case these are W and Z
bosons, top and stop quarks and h and H bosons with multiplicities:
\begin{equation}
g_w=6~,~g_z=3~,~g_t=g_{\tilde{t}}=12~,~g_h=g_H=1
\end{equation}

We have neglected the contributions due to other quark-squark flavors.
This is justifiable insofar as their masses are small. As was
pointed out in ref.[7], also the bottom-sbottom contributions can be
non negligible for very large values of $tan\beta $. This case
will not be relevant for us.

\section{Nonzero Temperature Effective Potential}

When the temperature is nonzero, effective potential gets a
contribution [19]:
\begin{equation}
\triangle V_T=\frac{T^4}{2 \pi^2} \sum_i g_i I_{\pm}
    \left[ \frac{m_i(\phi)}{T} \right]
\end{equation}
where $g_i$ are multiplicities as before, $m_i(\phi )$ are field
dependent masses and $I_-(I_+)$ which are to be used for bosons
(fermions) are given by:
\begin{equation}
I_{\pm}(y)=\mp \int_0^\infty x^2 \ln \left( 1\pm e^
        {-\sqrt{x^2+y^2}} \right) dx
\end{equation}
This contribution to the effective potential describes the interactions
of the Higgs bosons with the thermal bath surrounding them.
Expressions (22) are rather difficult to operate with, especially when
the masses depend on fields in a complicated way.
Fortunately, as shown in ref. [2], one can always use either high
temperature (small y) or low temperature expansion (high y), so that the
mistake in determining $\triangle V_T$ is never bigger than 10 percent.

High temperature expansions of (4.2) are given by:
\begin{eqnarray}
h_-(y)=-\frac{\pi^4}{45}+\frac{\pi^2}{12}y^2 &-&\frac{\pi}{6}
       y^3-\frac{y^4}{32}\ln \left( \frac{y^2}{c_b}\right),
h_+(y)=-\frac{7 \pi^4}{360}+\frac{\pi^2}{24}y^2+\frac{y^4}
       {32} \ln \left(\frac{y^2}{c_f}\right),\nonumber \\
& &\ln c_b \approx 5.41,\ln c_f \approx 2.64,
\end{eqnarray}

Whereas the low temperature expansion is:
\begin{equation}
l(y)=-\sqrt{\frac{\pi}{2}}y^{3/2}e^{-y}
     \left(1+\frac{15}{8 y} \right)
\end{equation}

In this calculation we will always use one of these expansions or
linear interpolation between them.
By substituting field dependent values of $m_w$, $m_z$,
$m_t$, $m_{\tilde{t} }$, $m_h$ and $m_H$ from
equations (13) - (17) into the expressions for the
effective potential (18) and (21) one obtains the full zero and
nonzero temperature one-loop effective potentials. The critical
temperature in this system is close to the point where the
temperature dependent effective mass matrix has a zero eigenvalue.

What can we get out of this? If we take $m_t$ and $\mu $ to be our
input parameters we have: 3 SUSY parameters $m_1$, $m_2$ and $m_3$;
2 zero-temperature VEV's $<\phi_1>$ and $<\phi_2>$,
2 nonzero-temperature VEV's  $<\phi_1>_T$ and  $<\phi_2>_T$
and the critical temperature -- altogether eight unknowns.

How many conditions do we have? First the fixed magnitude of the
zero T VEV (11) and (12),
then 2 zero-T minima:
\begin{equation}
\frac{\partial V^0}{\partial \phi_1}=
\frac{\partial V^0}{\partial \phi_2}=0
\end{equation}
next 2 nonzero-T minima:
\begin{equation}
\frac{\partial V^T}{\partial \phi_1}=
\frac{\partial V^T}{\partial \phi_2}=0
\end{equation}
the critical temperature condition:
\begin{equation}
\det \left( \frac {\partial^2 V^T}
{\partial \phi_i \partial \phi_j} \right)_{\phi_1=\phi_2=0,T_c}
\approx 0
\end{equation}
and, finally, from (1) and (2), the condition that nonzero-T VEV
be sufficiently large:
\begin{equation}
v(T_c)=\sqrt{v^2_1(T_c)+v^2_2(T_c)}\geq v_{crit}(T_c)=
        \frac{45 g_w T_c}{4 \pi B(\lambda/g^2_w)}
\end{equation}

If we take the equality in (28) (which corresponds to the upper limit on
the $m_h$ ), we have seven equations in eight unknowns, therefore we can
impose one relation between them. This was done numerically in the form:
$m_h=m_h(tan\beta )$ for different values of parameters $m_t$ and $\mu $.
Here, $m_h$ is the upper limit on the mass of the lighter CP-even Higgs
field.

This program was realized in MATHEMATICA and
maxima of curves $m_h(tan\beta)$ are given in
figures 1,2 and 3 for $m_t$ = 115 GeV, 150 GeV and 200 GeV respectively.

The argument $\lambda^{eff}/g^2$ in the function $B(\lambda^{eff}/g^2)$
was determined in ref. [5] for the general case of a two doublet model:
\begin{equation}
\lambda^{eff}=\lambda_1 \cos^4 \beta_T+\lambda_2 \sin^4 \beta_T+
2 h \cos^2 \beta_T \sin^2 \beta_T
\end{equation}
Here $\beta _T$ is the nonzero-T ``mixing angle of VEV's'':
\begin{equation}
\tan\beta_T=\frac{v_2(T_c)}{v_1(T_c)}
\end{equation}
In the MSUSY case:
\begin{equation}
\lambda_1=\lambda_2=\frac{g^2_1+g^2_2}{4},h=-\frac{g^2_1+g^2_2}{4}
\end{equation}
therefore:
\begin{equation}
\lambda^{eff}=\frac{g^2_1+g^2_2}{4}\cos^2(2 \beta_T)
\end{equation}

There are several causes of uncertainty in this calculation.
First, we have assumed that the phase transition happens at point
$T_c$ where (27) is satisfied. This is not true. It was already
noticed in ref. [1] that since the phase transition happens
earlier, when the vacuum expectation value is smaller than at
$T_c$, the actual bound is stronger than the one we take.
In other words, if we were able to calculate the phase transition
temperature exactly the upper limit on the Higgs mass would be
lower than the one we impose. Unfortunately, at the phase
transition the one loop effective potential is not accurate at
the origin due to infrared divergences and therefore we can just
estimate the critical temperature.

Secondly, when using eq. (28) for the MSUSY we used the fact
derived in [5] that the upper bound on the sphaleron mass in
MSUSY is the sphaleron mass of the Standard Model. Therefore, our
bound is again weaker than the actual one but still it will turn
out to be very strong. However, the more precise calculation
would require calculating the sphaleron mass in MSUSY at the
critical temperature.

Finally, as is usual in the study of phase transitions in early universe,
we are using effective potential which is a static
quantity for a system which not only evolves but evolves out of
equilibrium.

One should keep all of these caveats in mind when interpreting
the results which are given in the next section.

\section{Results and Discussion}

Figures 1, 2 and 3 show upper limits on the Higgs mass for
mt=115, 150 and 200 GeV respectively. Points are obtained as
maxima of curves $m_h(tan\beta)$ for different values of $\mu$.
For easier visibility they have been connected by straight lines.
One can draw two
conclusions from these results.

First, for values of $\mu=150$GeV (which is the asymptotic
lower experimental mass limit at 90\% c.l. for a gluino mass
lower than 400 GeV - see ref. [20] ), we get the {\em
upper limit on the Higgs mass to be 51 GeV, 54.5 GeV and 63 GeV
for 3 different values of the top mass.} One should compare this
with {\em experimental lower limit of 41 GeV }\footnote{This limit does
not apply when $m_A < 2 m_{\mu}$, where A is the CP-odd Higgs
boson.} (see [14] and [21] ).

Second, for the considered region of $tan\beta > 1$ ,
there was always a maximal value of $\mu$ above which $v(T_c)$
was never big enough to satisfy requirement (28). This gives
{\em the upper limit on the soft-SUSY breaking scale of about
750 GeV, 250 GeV and 170 GeV for three top masses considered.}
One should compare this with the previously mentioned asymptotic
{\em lower mass limit of 150 GeV.}

As a conclusion one can establish the following ``no loose
theorem'' from these results:
{\em either} the top quark is heavy (fig 3.) in which
case Higgs can be as heavy as 63 GeV but SUSY breaking scale is
very close to its experimental lower limit {\em or} top is
lighter than 150 GeV (fig 1. and 2.) but then the Higgs mass
is lighter than 55 Gev and thus close to its experimental lower
limit.

As with all other calculations in supersymmetric models this one has
the trouble that there are simply too many unknown parameters.
We have considered that region in parameter space which has been
searched by experiments [14] ,[20], [21] ( i.e. for a common soft
supersymmetry-breaking mass for $\tilde{t}_L$ and $\tilde{t}_R$
and vanishing off-diagonal mass elements for the stop mass matrix ).
The limits that we obtain complement the experimental results and
severely limit the parameter space of MSUSY.

{\bf Acknowledgments} This research was supported by grants from
NSF, DOE and Texas National Research Laboratory Commission.

I would like to thank Mitchell Golden and Stephen Selipsky for useful
discussions, Christian Mannes for help with MATHEMATICA and
particularly my advisor Andrew Cohen for proposing this project
to me and for discussing it at all its stages.

\end{document}